\documentclass[twocolumn]{revtex4}
\usepackage{amssymb,amsfonts,amsmath}
\usepackage{color}
\usepackage{graphicx}
\usepackage{multirow}
\usepackage{longtable}
\usepackage{setspace}
\usepackage{epsfig, epstopdf,latexsym,graphicx,bm,pifont}

\providecommand\bdelt{\mbox{\boldmath$\delta$}}

\def\kt{k_\text{B}T}

\def\Lee{L_\text{ee}}
\def\fp{f_\text{p}}
\def\fc{f_\text{c}}
\newcommand{\vect}[1]{\mathbf{#1}}
\newcommand{\vhat}[1]{\mathbf{\hat{#1}}}
\def\bfp{\vect{f}_\text{p}}
\def\ka{\kappa_{\text{a}}}
\begin{document}

\begin{titlepage}
\title {Flagellar dynamics of a connected chain of active, Brownian particles}
\author{Raghunath Chelakkot$^1$, Arvind Gopinath$^1$, L. Mahadevan$^{2,3
*}$, Michael F. Hagan$^1$}
\email{hagan@brandeis.edu,
lm@seas.harvard.edu}
\affiliation{$^1$Martin Fisher school of physics, Brandeis University,
Waltham, MA 02453, USA\\
$^2$School of Engineering and Applied Sciences, Harvard University, Cambridge, MA 02138,  USA\\
$^3$Department of Physics, Harvard University, Cambridge, MA 02138, USA
}

\begin{abstract}

Eukaryotic flagella are active structures with a  complex architecture of microtubules, motor proteins and elastic links. They are capable of whiplike motions driven by motors sliding along filaments that are themselves constrained at an end. Here, we show that  active, self-propelled particles that are connected together to form a single chain that is anchored at one end can produce the graceful beating motions of flagella. We use a combination of numerical simulations, scaling analysis and mean field continuum elastic theory to demarcate the phase diagram for this type of oscillatory motion as a function of the filament length, passive elasticity, propulsion force and longitudinal persistence of propulsion directions. Depending on the nature of the anchoring, we show that filament either undergoes flagella-like beating or assumes a steadily rotating coiled conformation. Our system is simpler than its biological inspiration, and thus could be experimentally realized using a variety of self-propelled particles.
\end{abstract}
\maketitle
\end{titlepage}
\begin{figure}
\includegraphics[width=\columnwidth]{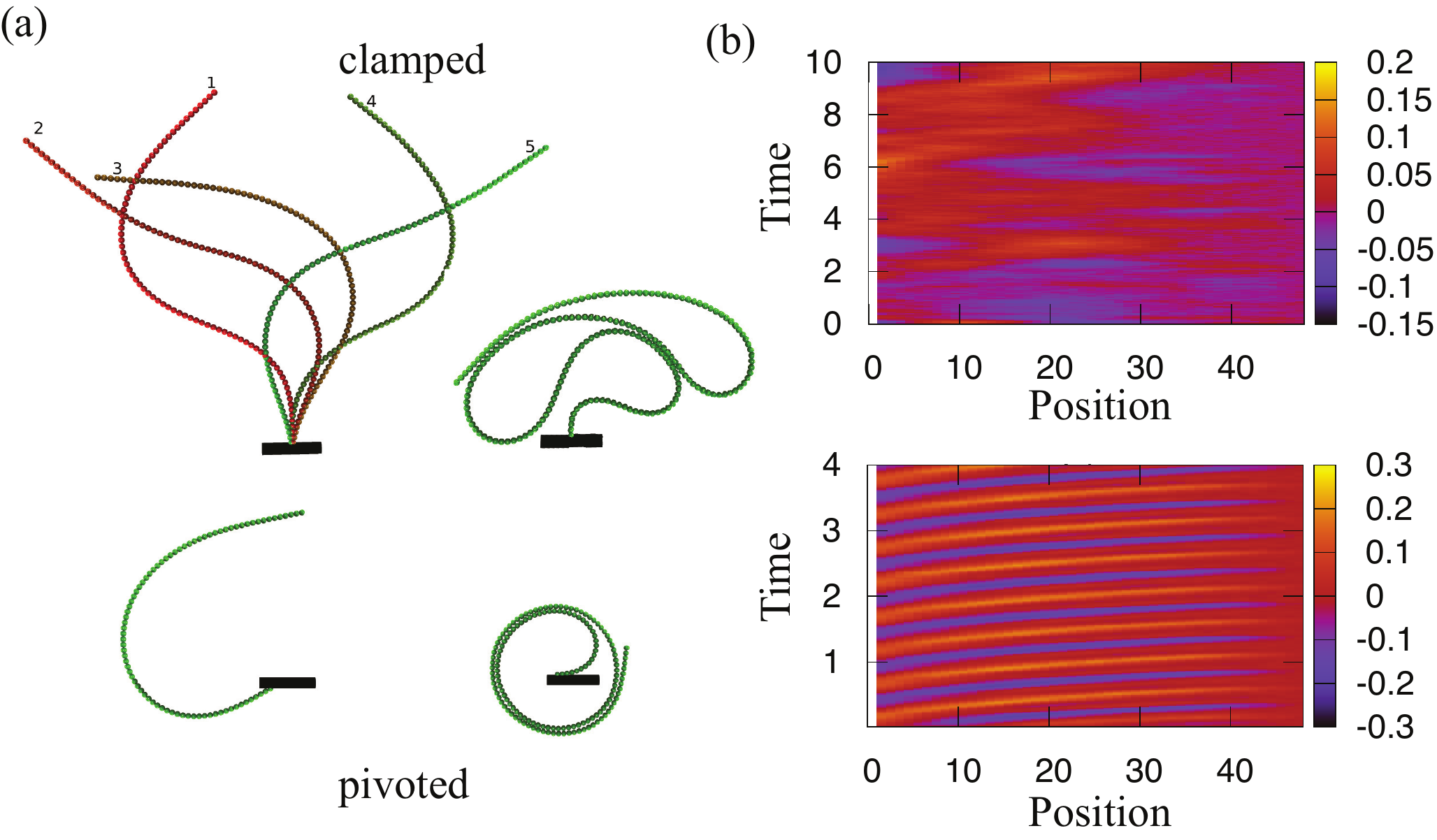}
\caption{(a) (Clamped end) Left: Shapes taken by a stably oscillating short filament over one beat cycle - the slight asymmetry is a consequence of noise. Right: A self-contacting, stably beating, long filament. (Pivoted end):  Left: A stably rotating filament for small lengths similar to those observed in motility assays \cite{Bourdieu1995}.  Right: A tightly coiled shape for large lengths. (b) Filament curvature as a function of time and the distance from the clamped end for (top) non-periodic beating, and
(bottom) periodic buckling. Simulation movies are in the supplementary information~\cite{movies}}
\label{fig1}
\end{figure}
Eukaryotic cilia and flagella are whiplike, elastic micro structures that undergo oscillatory beating to drive processes such as locomotion~\cite{Bray}, mucus pumping~\cite{Fulford}, embryogenesis~\cite{Nonaka}, and directed cell migration \cite{Sawamoto}. While
the molecular mechanisms that control ciliary beating remain incompletely understood, it is well established that sliding forces generated by dynein motors attached to the microtubule-based backbone of cilia play a crucial role ~\cite{flagella}. In addition to understanding how these active structures work in nature, there is growing interest in designing artificial analogs.
Recent experiments on a minimal motor-microtubule system \cite{Sanchez} demonstrate cilia-like beating, and artificial beating systems driven by external periodically varying electromagnetic fields have been synthesized ~\cite{artificial_cilia}. However,  internally driven structures capable of controllable beating patterns have yet to be developed.
In this letter we use simulations and theory to identify a different mechanism that results in controllable, internally driven flagella-like beating or steady rotation, in microstructures comprised of connected self propelled units. The resulting shapes and motions are a consequence of a fundamental instability of a filament undergoing tangentially-directed compressive forces - also termed {\em follower forces}.

Our model filament is a two-dimensional chain of connected, self-propelled colloidal spheres. They may be experimentally realized, for example, as Janus beads in varying contexts ~\cite{diffusophoresis, thermophoresis,granular} and can potentially be connected to form active filaments, either through face-face attractions or by passive tethers\cite{Vutukuri2012agnewchemie,Sung2008,Sacanna2012,Vutukuri2012advMater}. Here we assume that the filament has $N$ self-propelled polar spheres each of diameter
$\sigma$ located at coordinates
$\vect{r}_i$ ($i=1,..,N$). The spheres are connected by harmonic potentials of equilibrium length $b$, $U_\text{l}=\frac{\kappa_\text{l}}{2}\sum_{i=1}^{N-1} \left( |\vect{r}_{i+1}
-\vect{r}_{i}| -b \right)^2$,  and resistance to bending is implemented via a three-body bending potential
$U_\text{b} =\frac{\kappa}{2}\sum_{i=2}^{N-1} \left(\Hat{b}_{i+1}
-\Hat{b}_{i}  \right)^2$,
where $\Hat{b}_i =(\vect{r}_i-\vect{r}_{i-1})/|\vect{r}_i-\vect{r}_{i-1}|$ is the unit bond vector and
$\kappa$ is the bending rigidity.
Each sphere exerts a local force on the filament due to self-propulsion. This active force
$\bfp^i$ of constant magnitude acts on each sphere
along its polarity vector $\vect{p}_i$, such that $\bfp^i=\fp \vect{p}_i$ with $|\vect{p}_{i}| = 1$. We consider self-linked Janus particles with a homogeneous polarity along the filament, i.e. the polarity vector itself tends to point along the linking (tangent) direction $\vect{b}_i$ under the influence of a harmonic potential $U_\text{a} =\frac{\ka}{2} ( \vect{p}_i- \vect{b}_i)^2$.
 For finite $\ka$, the angular potential biases the propulsion forces to act along  the local tangent vector of the filament; in the limit $\ka \rightarrow \infty$, the propulsion forces  are perfectly aligned with the local tangent.  The arc-length parameter $s$ in the range $0 \leq s \leq \ell \equiv Nb$ parameterizes the coarse-grained position along the filament.

The dynamics of the filament is determined by the evolution of the sphere positions $\vect{r}_i$ and orientations
$\vect{p}_i$, which we simulate using overdamped Brownian Dynamics $(i=1,..,N)$,
\begin{multline}
\begin{cases}
\dot{\vect{r}}_i=\beta D\:(\vect{F}_i^\text{l}+\vect{F}_i^\text{b}+\vect{F}_i^{\mathrm{Ex}}+\fp{\vect{p}})+\sqrt{2D}\:{\boldsymbol\zeta}_i, \\
\dot{\vect{p}}=\beta D_{\mathrm{r}} \:\vect{F}_\text{a} (\vect{p}_i)+\sqrt{2D_{\mathrm{r}}} \:
{\boldsymbol \zeta}_i^\text{R}.
\end{cases}
\end{multline}
Here $\vect{F}_i^\text{l}=-\partial{U_\text{l}}/\partial{\bf r_i}$, $\vect{F}_i^\text{b}=-\partial{U_\text{b}}/\partial{\bf r_i}$, and $\fp \vect{p}$ are respectively the bond, bending, and propulsion forces, $\vect{F}_\text{a}$ is the angular force derived from the angular potential $U_\text{a}$, and $\vect{F}_i^{\mathrm{Ex}}=-\partial{U_\text{ex}/\partial{\bf r_i}}$ is an excluded-volume repulsive
force given by a truncated and shifted Lennard-Jones potential for interparticle distance $r$: $U_{\mathrm{ex}} = 4 \epsilon \left[
  \left( \frac{\sigma}{r} \right)^{12} - \left( \frac{\sigma}{r}
  \right)^6 \right] + \epsilon$ if $r < 2^{\frac{1}{6}}$, and zero
otherwise, and $\beta=1/\kt$. $D$ and $D_{\mathrm{r}}$ are
translational and rotational diffusion constants, which in our low-Reynolds-number regime are set to satisfy
the Stokes-Einstein relationship  $D_{\mathrm{r}}=
{3D}/{\sigma^2}$. $\boldsymbol \zeta$ and

${\boldsymbol \zeta}^\text{R}$ are unit variance Gaussian white noise forces
and torques respectively, each of which satisfies
$\langle {\boldsymbol \zeta}_i \rangle=0$ and
$\langle {\boldsymbol \zeta}_{i}(t) {\boldsymbol \zeta}_{j}
(t^\prime )\rangle = 2\delta (t-t' ){\bdelt}_{ij}$.
As the viscous mobility of a sphere is isotropic, our simulations neglect differences in viscous resistance to normal and tangential modes of filament motion.  However,  simulations including  full hydrodynamic interactions, using  a hybrid simulation technique~\cite{mpc},  display very similar qualitative features  (see SI $\S$III). We make the equations dimensionless by using $\sigma$ and $\kt$ as basic units of length and energy, and ${\sigma^2}/{D}$ as the unit of time, and we set $\epsilon=\kt$.

For our numerical simulations, runs were initialized using a straight configuration with the filament vertically oriented along $\vhat{e}_x$, with all ${\vect{p}}_i$ initially along
$-\vhat{e}_x$. One end of the filament corresponding to $s=\ell$ was always free. The anchored end $s=0$ was either clamped vertically
(Fig~\ref{fig1}(a) top) ~\cite{footnote1} or attached to a frictionless pivot (Fig~\ref{fig1}(a) bottom) with the filament free to rotate.
\begin{figure}
\includegraphics[width=\columnwidth]{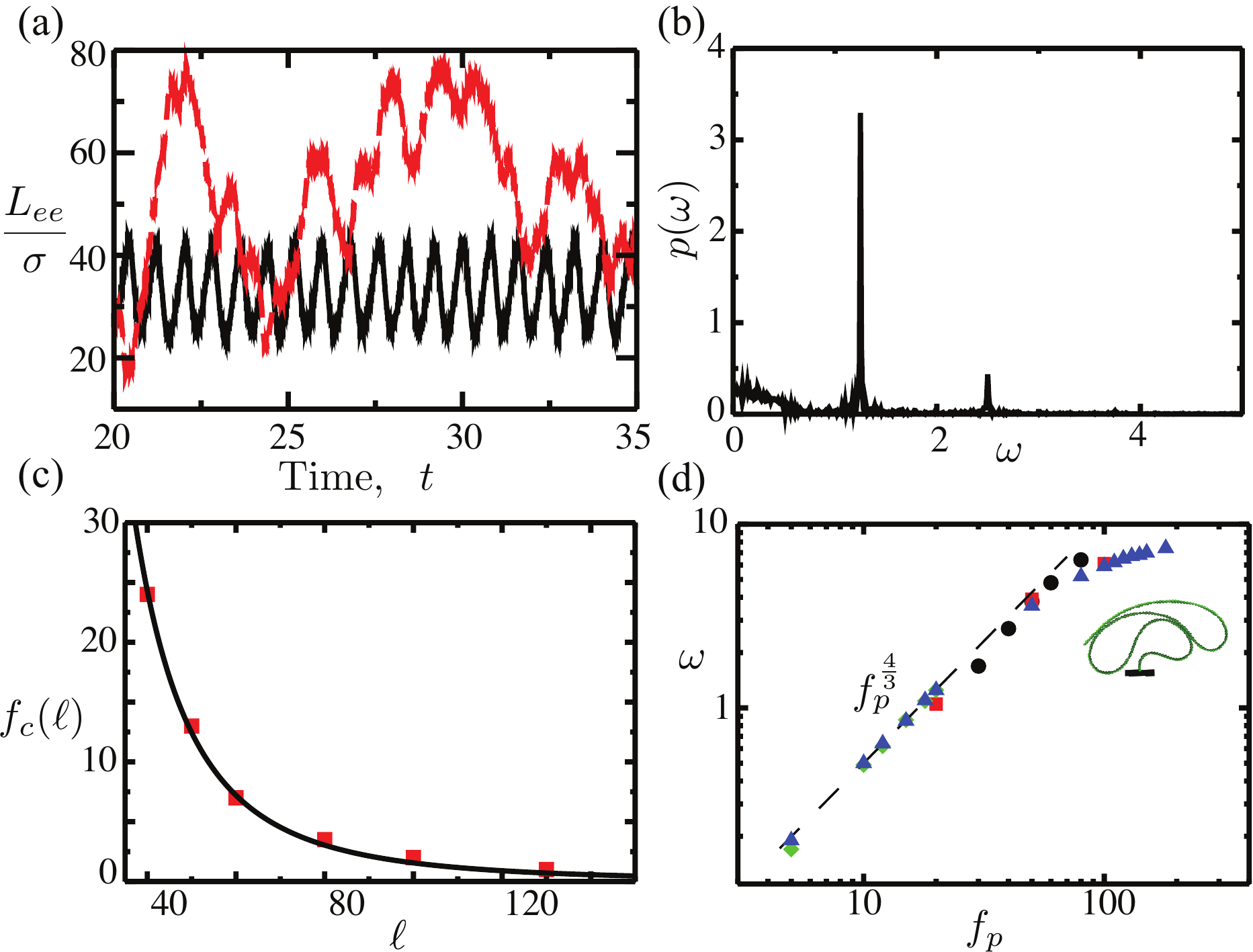}
\caption{(a) Length of the end-end vector $\Lee$ for periodic beating (full line) and non-periodic buckling (broken line), (b) Power spectral density of the end-end distance for filament length $\ell=80$, propulsion force $\fp = 20$, and angular force constant $\ka = 20$,
(c) critical propulsion force $\fc(\ell)$ for various filament lengths, at angular force constant $\ka = 100$. The solid line is $\fc=C\kappa/\ell^3$ with $C\simeq78$, consistent with (\ref{Eq:2}). (d) The beating frequency of filaments against propulsion force for filament lengths $\ell  = 40 \:(\bullet)$, $\ell  = 50 \:(\blacksquare)$, $\ell = 80 \:(\blacklozenge)$, and $\ell = 100 \:(\blacktriangle)$. The dashed line corresponds to the scaling law(\ref{eq:3}). Parameters are dimensionless as described in the text. For very long lengths, the filament starts to self-contact (inset) resulting in a decrease in the exponent due to contact friction. The value of bending stiffness is $\kappa=2\times10^{4}$ in all figures.}
\label{fig2}
\end{figure}

{\it Results for a clamped filament:} For the clamped case ${\bf b}_0= \vhat{e}_x$, and ${\bf r}_0=0$. When the internal propulsion force $\fp$ exceeds a critical value $\fc(\ell,\ka)$, the filament buckles, in a manner similar to a self-loaded elastic filament subject to  gravity \cite{gravity}. However the post-buckled states are quite different (see SI-$\S$I) since the direction of active force density follows polarity vectors $\vect{p}_i$ and thus tends to point along the filament axis.

At fixed filament stiffness $\kappa$, and for $\fp>\fc(\ell,\ka)$, we find that the magnitude of the polarization stiffness $\ka$ controls the long time dynamics of the filament. In figure~\ref{fig1}(b - top) we show the local filament curvature as a function of time and arc length. For small values of
$\ka$, thermal diffusion controls the local orientation of $\vect{p}_i$ and the propulsion activity is uncorrelated along the filament. In this regime, the filament dynamics is marked by transients resulting from the bending generated by the particle propulsion, but no coherent patterns. When $\ka$ is increased above a critical value $\ka^\text{c}$, the polarities of the spheres align strongly with the local tangent. The resultant self-propulsion force is strongly correlated with the filament tangent and the filament oscillates with periodic, large amplitude wavy motions shown in Fig~\ref{fig1}(b - bottom) that propagate from the proximal (clamped) to the distal (free) end. This beating profile is very similar to that of flagella in eukaryotic cells though the underlying physics differs fundamentally, since there is just a single filament made of active particles here.

We quantify the regularity of these oscillations by measuring the length of the end-end vector $\Lee =\sqrt{(\vect{r}_N-\vect{r}_1).(\vect{r}_N-\vect{r}_1)}$, as a function of time, shown for  two values of $\ka$, $\ka<\ka^\text{c}$ and $\ka \gg \ka^c$, in Fig.~\ref{fig2}a. While $\Lee$ displays large variations in time for both values of $\ka$ because of the large propulsion force, the profile for $\ka \gg \ka^\text{c}$ is periodic, and its power spectral density (PSD) shows a distinct frequency maximum (Fig.~\ref{fig2}b).

To understand our numerical experiments and obtain estimates for the critical buckling load and the frequency of ensuing oscillations in the periodic regime, we consider the limit $\ka \rightarrow \infty$, and coarse-grain the chain of spheres into a slender, elastic filament of  length $\ell$ and bending stiffness $\kappa$. The force due to the self-propulsion translates to a compressive force per unit length of strength $\fp$ oriented anti-parallel to the local tangent vector. For the clamped filament, the straight filament becomes unstable to time-dependent shapes beyond a critical value of $\fp$, when the resultant internal propulsion force $\fp \ell$ deflects the tip by a small transverse distance $h \ll \ell$ leading to an effective filament curvature $O( h/\ell^2)$. Balancing moments about the base then yields $
h \: \fp \ell  \sim \kappa {h / \ell^2} $
and thence the critical force beyond which the straight filament is no longer stable,
\begin{equation}
\fc \sim \kappa / \ell^3.
\label{Eq:2}
\end{equation}
This scaling relationship closely agrees with computationally determined values (Fig~\ref{fig2}c). To determine the oscillation frequency we note that when the filament buckles, it is always under compression and no small amplitude, stable steady solution exists (SI-$\S$1). When $\fp \gg \fc$, the characteristic length over which the active compression is accommodated  scales as $\lambda   \sim (\kappa / \fp)^{1 \over 3} \ll \ell$. In the over-damped limit, all the energy supplied by the self-propulsion transforms first into elastic bending energy before being ultimately dissipated viscously. In a time $\omega^{-1}$, the energy dissipated viscously is the product of the force per unit length $\eta_{\perp} \lambda \omega$, the characteristic deflection $\lambda$, and the velocity $\omega \lambda$. This dissipation has to balance the active energy input into the system due to the self propulsion $\fp \lambda^{2} \omega$ and thus
$
\eta_{\perp} \lambda^3 \omega^2 \sim \fp \lambda^{2} \omega.
$
Using our previous expression for $\lambda \sim (\kappa / \fp)^{1 \over 3}$, we obtain
\begin{equation}
\omega \sim {\eta_{\perp}}^{-1} \left({\fp^{4} / \kappa}\right)^{1 \over 3}
\label{eq:3}
\end{equation}
which is also
in excellent agreement with our simulations (Fig.~\ref{fig2}d). A systematic derivation of the equations governing the filament dynamics  that builds on a local resistivity formulation \cite{Brenner} relating  filament bending to its velocity yields two coupled non-linear equations for the tension in the filament $T$ and $\theta$, the angle the filament makes with $\vhat{e}_{x}$ (see SI-$\S$2 for details) and confirms this simple scaling picture.

{\it Results for a filament anchored at a pivot.} The beating discussed thus far is a consequence of the clamped boundary condition which prevents rotation as well as translation. We next
performed simulations for a filament moving about a frictionless, pivoting end at $s=0$ such that ${\bf r}_0=0$ whereas ${\bf b}_0$ is unconstrained.

For small values of
$\ka$, and with the contour length $\ell$ and rigidity $\kappa$ held fixed, the filament end-end length $\Lee$ displays large irregular variations and the end-end vector $\vect{L}_\text{ee}$ undergoes irregular rotation about the fixed end.
Increasing $\ka$ results in the active forces being increasingly correlated along the contour. The post-buckled filament now assumes a steadily rotating bent shape and the value of $\Lee$ does not vary in time (Figure 1a).

\begin{figure}
\includegraphics[width=\columnwidth]{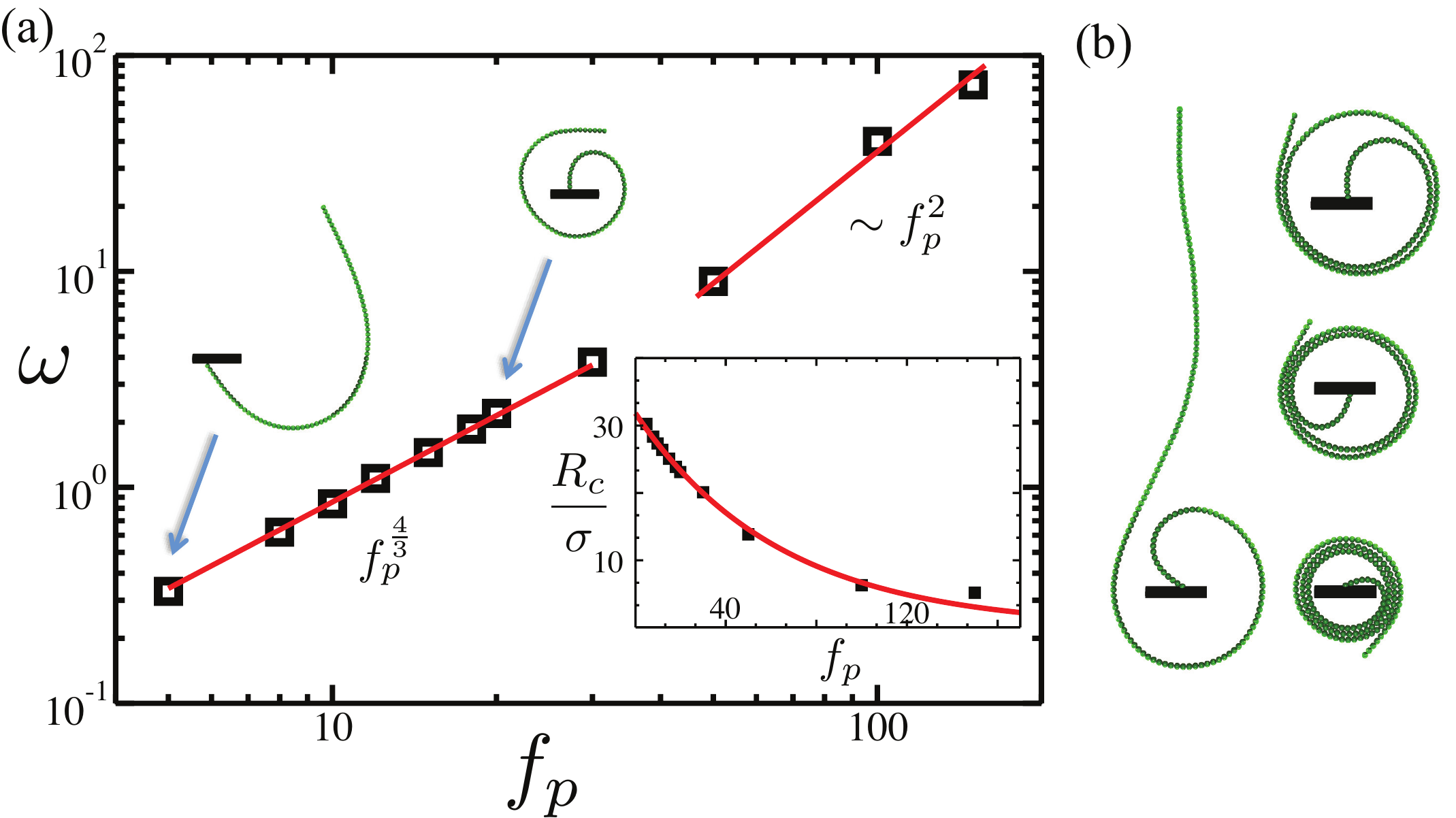}
\caption{(a) Frequency of steady rotation for a filament anchored at pivot, with length $\ell=160$, angular force constant $\ka=2\times 10^{2}$, and bending stiffness $\kappa = 2\times 10^{4}$. The inset shows the radius of the coiled shape as function of $\fp$. For very large forces, long filaments yield rotating  closed coils and the frequency scales differently due to the excluded volume constraint. (b) Transient shapes for very large $f_{p}$ as the filament coils and eventually undergoes steady rotation.}
\label{fig3}
\end{figure}

The rotation frequency extracted from simulations by calculating the orientation of $\vect{L}_\text{ee}$ as a function of time is plotted in Fig~\ref{fig3} as a function of $\fp$. For $\ell \ll 3 \pi (\kappa /\fp)^{1 \over 3}$, the wavelength of the bend is larger than the length of the filament and the steady rotational frequency observed varies with force in accord with equation (3). The shapes we obtain and the dependence of the frequency on the active force compare well with the experimental observations of motility assays where filaments animated by underlying molecular motors encounter pinning sites (defects) and start to
rotate \cite{Bourdieu1995}.
Beyond a critical value of $\fp$ (for  $\ell$ fixed)  or when
$\ell \gg 3 \pi (\kappa /\fp)^{1 \over 3}$ (for $\fp$ fixed), the wavelength of the bend becomes smaller than the length of the filament thus
enabling self-contact.  Since excluded volume interactions prevent overlap,  the filament assumes increasingly
 coiled shapes as shown in Fig~\ref{fig1}a and in Fig~\ref{fig3}a.
Once coiled, the frequency of steady rotation depends strongly on friction due to relative sliding and excluded volume forces. Our simulations indicate that for fixed $\ell$, the frequency almost scales as $f^{2}_{p}$ (Fig 3a). Relaxing the excluded volume constraint results in completely overlapping coils and a frequency that scales as $\fp^{4/3}$, consistent with (\ref{eq:3}).

Snapshots of the transient rotating shapes seen for long filaments (Fig 3b) show evidence of multiple time scales in the coiling process. Initially, the filament  rotates and and adjusts its curvature until the free end almost touches itself. At this point, the remaining length is accommodated in adjacently placed coils. Finally, as steady rotation is attained, the effective radius of the innermost coil $R_{c}$ reduces gradually, the coils tighten and the number of coils increases
slightly. The radius $R_{c}$ decreases very sharply with $\fp$ before eventually bending of the coil is balanced by the lateral forces due to excluded volume, resulting in a very slow decay with  $\fp$ (Fig 3a inset).

\begin{figure}
\includegraphics[width=\columnwidth]{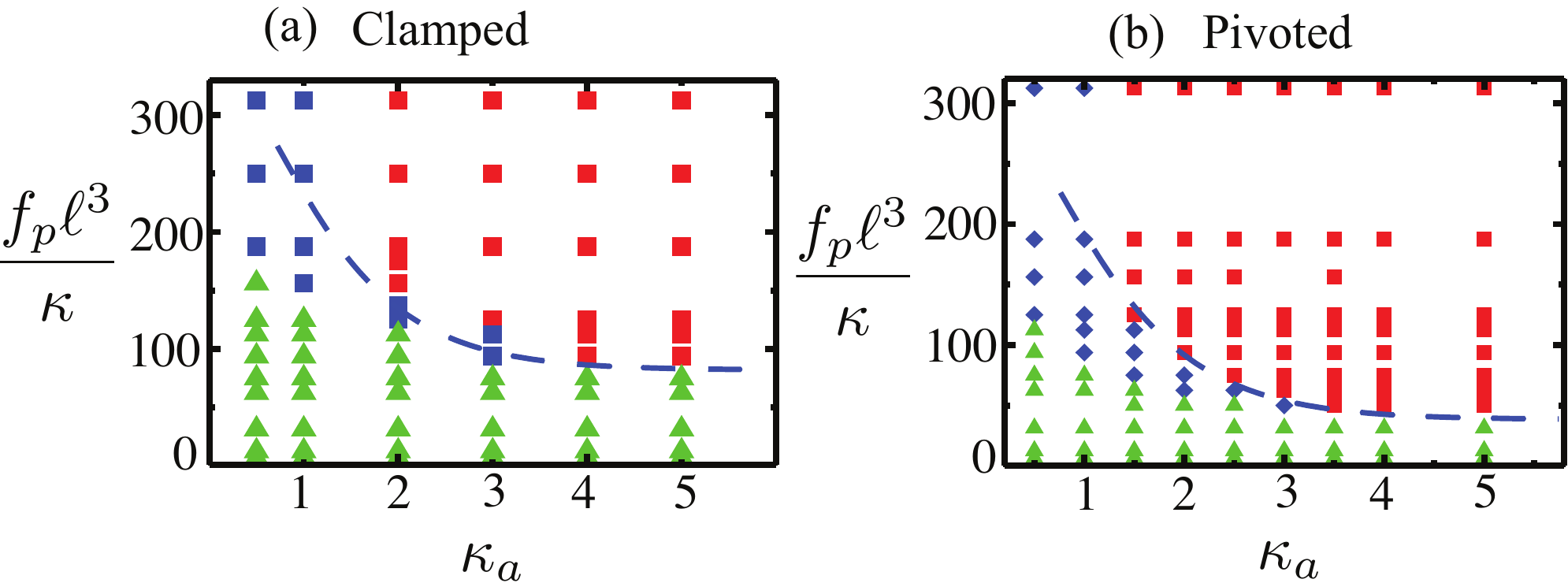}
\caption{Phase diagram for (a) a clamped filament
and (b) a rotating filament
with bending stiffness $\kappa=2\times10^{4}$. Green triangles correspond to no beating, blue symbols to irregular beating, and red squares to regular beating. The blue dashed curve (valid for $\ka \gg 1$) corresponds to equation (4).
The constants were determined by eye and are (a) $(A,\:B)  = (85, 80) $ and (b) $(A,\:B)  = (36, 70)$.}
\label{fig4}
\end{figure}

{\it Dynamical correlations along filament:} Having delineated the different dynamical regimes of the active filament as a function of the active force and the anchoring condition, we now investigate the dynamical behavior of the connected self-propelled particles as a function of the angular stiffness $\ka$ and self-propulsion force density $\fp$. In Fig~\ref{fig4}, we show the phase diagrams for both clamped and pivoted filaments.

For propulsion forces below the critical value $\fp < \fc$, we do not  observe  any statistically significant beating (green triangles) - the straight configuration remains stable and variations in $\langle \Lee\rangle\simeq \ell$ due to fluctuations in the directions
of $\vect{p}_{i}$ are insignificant. The critical value of $\fc$ that separates this regime from the unsteady (periodic as well as non-periodic) regime is a function of the passive filament elasticity ($\ell$ and $\kappa$) and also of $\ka$.
Keeping $\ka$ fixed and increasing $\fp$ to beyond the critical value renders the  straight configuration unstable and results in stable unsteady shapes.  For small values of $\ka$ ($\leq 1$) we observe irregular beating (Fig.~\ref{fig1}b, top).
As  $\ka$ is increased ($\ka > 1$), enhanced correlation of particle propulsion results in two trends - evident both for the clamped (Fig 4a) and pivoted (Fig 4b) cases. First, the critical force to destabilize the straight configuration decreases, saturating as $\kappa \rightarrow \infty$ to the continuum limit, consistent with  (2). The increasing correlation between the self-propulsion directions of the spheres  results in periodic beating (red squares, Fig.~\ref{fig1}b, bottom).
Interestingly however, at any $\ka \gtrsim 1$, there persists a diminishing region of irregular beating between the no beating and the periodic beating regions. This behavior arises when  the correlation time of  propulsion directions is shorter than the filament oscillatory timescale in the perfectly aligned limit.

To understand the location of the boundary between regular and irregular beating, we note that regular oscillations will be interrupted when a critical number of neighboring spheres escape from an configuration driving the beating to a  configuration that is not unstable - chosen here to be the configuration with the polarity vector along the local tangent (and thus the force being extensional). The mean time  required for this transition to occur $T_{c}$ may be simply estimated using a classical (mean) first-passage time analysis \cite{Hanggi}. The angular potential in terms of the angle $\zeta$ between ${\bf{p}}$ and the tangent vector is $U_{a} = \ka (1-\cos{\zeta})$. The base of the potential corresponds to $\zeta = 0$ and the energy when $\zeta = \pi$ is $2\ka$ (in $k_{b}T$ units). For $\ka > 1$, equilibration in the angular escape coordinate $\zeta$ may be assumed with transitions occurring diffusively. Using the expression for the frequency of crossings in the over-damped limit (SI - \S 3), we find that the critical curve satisfies
\begin{equation}
({\fp \ell^3 \over \kappa}) \sim A + B \: \ka^{3 \over 4} (e^{-2\ka})^{3 \over 4}.
\end{equation}
Here $A$ is a constant that ensures that the critical force matches the continuum limit ($\ka \rightarrow \infty$), and  $B$ is to be interpreted as the minimal size of the patch (equivalently force) required to cause beating irregularities.  While strictly valid for $\ka \gg 1$, this expression - plotted as the dashed blue line in Figs. 4a and 4b - satisfactorily captures the shape of the boundary for $\ka > 1$.

\emph{Conclusions.}  Our study suggests a simple proposal to mimic the beating of eukaryotic flagella. Rather than having motors that walk on adjacent filaments that are clamped at an end, we have shown that we can generate actively oscillating and rotating filaments using connected self-propelled particles. The frequency of oscillations and thereby the swimming speed and fluid forces can be controlled by varying the dimensionless parameters in our problem - the ratio of the chain stiffness to the polarity stiffness $\kappa/\ka$, the scaled active force $\ell^3\fp/\kappa$, and the aspect ratio of the chain $\ell/\sigma$, or equivalently the number of active particles. Furthermore, if the angular potential is written as $U_\text{a} =\frac{\ka}{2} [( \vect{p}_i- \vect{b}_i)^2-C]$, where $0<C\leq1$, thus introducing a preferred direction different from the local tangent, the initial symmetry-breaking bifurcation is eliminated and the filament will always beat. All of these parameters may be accessible experimentally. 

Recent efforts to manipulate connected passive colloids by electrical fields \cite{Vutukuri2012agnewchemie}, microfluidics \cite{Sung2008}, lock-and-key type interactions \cite{Sacanna2012} and heat \cite{Vutukuri2012agnewchemie,Vutukuri2012advMater} have been successful in yielding externally actuated filaments with controllable stiffness. Extension of such techniques using diffusophoretic Janus particles as templates should yield internally controlled self-propelling filaments, just as a small variation of the motility assay for rotating filaments \cite{Bourdieu1995}, by clamping an end, will lead to beating. In a broader context, in contrast to most  studies of locomotion at low Reynolds number which prescribe the shape of the organism (typically as a slender filament with prescribed kinematics), here we prescribe the active forces locally, and calculate the resulting shapes. If the anchored end does not have infinite resistance the self-propelled particles will propel the whole chain, the study of which is a natural next step.

After our work approached completion, we learned of  a related numerical model~\cite{Jayaraman}.  However, our work differs in using a different formalism, different  boundary conditions, as well as a consideration of both scaling relations, and the underlying mechanisms.

{\bf Acknowledgements.} We acknowledge funding for this research provided by NSF-MRSEC-0820492, the MacArthur Foundation, and computational support from the Brandeis HPC. We thank Howard Stone for helpful comments on a preliminary version of this manuscript.


\begin{thebibliography} {31}
\bibitem{footnote1}
This is achieved by attaching the first particle at $(s=0)$ to a fixed point in space
$\vect{r}_0$ by a harmonic potential with the same properties as the binding potentials between the other particles in the filament. The three body-bending potential applied on
$\vect{r}_1$  has an equilibrium angle of $\pi/2$ between the bond vector $\vect{r}_1-\vect{r}_0$ and $\vhat{e}_y$.
\end{thebibliography}
\end{document}


\title {Flagellar dynamics of a connected chain of active, Brownian particles:\\ Supplementary Information}
\author{Raghunath Chelakkot$^1$, Arvind Gopinath$^1$, L Mahadevan$^{2,3}$ and Michael F. Hagan$^1$}
\affiliation{$^1$Martin Fisher school of physics, Brandeis University,
Waltham, MA 02453, USA\\
$^{2}$ SEAS, Harvard University, Cambridge MA 02138, USA.\\
$^{3}$ Department of Physics, Harvard University, Cambridge MA 02138, USA.
}

\date{\today}
\maketitle

\section{Comparison to the shape of a column buckling due to self weight}
In the continuum, noise-free  limit one may treat the array of connected active swimmers as a thin, elastic, inextensible filament - valid provided $\kappa b^2/k_{b}T \gg 1$ and $\sigma/\ell \ll 1$. The filament bends due to the action of compressive, active forces and the constraint of inextensibility.
Such an active filament, clamped vertically at one end and free at the other, remains stably vertical for small force (compression) densities.
At a critical value of the force density however, the straight shape is unstable to lateral perturbations and yields to a buckled shape.

Before buckling, the compressive forces all act vertically towards the clamped end and this scenario resembles the classical problem of a filament buckling under gravity. Originally proposed and solved by Euler in the 18th century \cite{Euler} this has been revisited again more recently by Keller and co-workers \cite{Keller} (also see references therein).  The correspondence arises due to the non-generic initial configuration with the  filament clamped vertically.
In the following, we will compare equations for this classical (passive) buckling problem  with those for an (active) self-propelled filament and demonstrate that  the post buckled states are very different.

Consider a cylindrical, slender column of length $\ell$, diameter $\sigma$, cross-sectional area $A = \pi \sigma^2 /4$, bending stiffness $\kappa$  and mass density $\rho$ clamped vertically upwards to a rigid flat surface at its base $s = 0$ and free at the other end $s =\ell$.  No external forces or torques act at the free end. We choose the origin of our Cartesian coordinate system to coincide with $s=0$, the $x$ axis to point vertically upwards and gravity to point downward such that $ {\bf g} =  - g {\bf e}_{x}$.
We parameterize the shape of the column by the angle, $\theta(s,t)$, the centerline makes with ${\bf e}_{x}$ and decompose the force resultant at a point $s$ along its length in terms of its Cartesian components,
${\bf F} = F_{x} {\bf e}_{x} + F_{y} {\bf e}_{y} $.
Torque and force balances on a elemental length yield (primes henceforth denoting differentiation with respect to arc-length, $s$)
$
{M''} = F_{x}\theta' \:\cos{\theta}\: +  F_{y} \theta' \:\sin{\theta}\: + A \rho g \:\sin{\theta}
$,
$
{F'_{x}} = A\rho g$, and ${F'_{y}}  = 0$.
Since there are no forces acting along the $y$ direction, we may integrate to find $F_{y} = 0$ and $F_{x} = A \rho g (s -\ell)$. Substituting this in the equation for the moment and using the definition $M = \kappa {\theta'} $, we find that the filament shape satisfies the differential equation
$
\kappa {\theta'''}  = A \rho g ((s-\ell) \sin{\theta})'
$.
In the small deformation limit $\sin{\theta} \approx \theta$, we scale length with $\ell$ and we obtain an equation for the angle $\theta$
\begin{equation}
{\theta'''} + \left( \alpha \:(1-s)\:\theta \right)' = 0, \:\:\:\:\:\alpha \equiv {\rho g A \ell^3 \over \kappa}
\end{equation}
with the associated boundary conditions $\theta(0) = 0$, ${\theta'} (1) = 0$ and ${\theta''} (1) =0$.

What is the corresponding equation for the filament of connected self propelled particles? Assuming that a static steady solution exists, we use a moment balance as
before. While it is easier to work in a local reference frame that uses the tangent and normal vectors at any point as the basis vectors, we choose to retain the Cartesian nomenclature to allow for comparison with equations for buckling due to gravity. The torque and force balances now yield
$
\kappa {\theta'''}  = {{\theta}'}(F_{x} \cos{\theta} + F_{y} \sin{\theta})$,
${F'_{x}} = f_{p} \cos{\theta}$ and  ${F'_{y}}  = f_{p} \sin{\theta}$.
Now making the small deformation assumption and using the fact that no external forces are present at the free end, we find that these equations reduce to
\begin{equation}
{\theta'''}  + \beta (1-s) {\theta'} = 0, \:\:\:\:\:\:\beta \equiv {f_{p}\ell^3 \over \kappa}
\end{equation}
where the constant $\beta \equiv (f_{p}\ell^{3}/\kappa)$ - a form different from equation (1).

The next question to answer is if steady small deformation solutions exist in either of these cases. We seek
non-trivial solutions in terms of the Airy functions ${\mathcal{A}}$ and ${\mathcal{B}}$.
Let us first consider the column buckling due to self-weight for which a solution
$
{\theta(s) \over C} =  {\mathcal{A}}(X) + {1 \over \sqrt{3}}\: {\mathcal{B}}(X)
$
exists with
constant $C$ undetermined at linear order and $X \equiv \alpha^{1 \over 3} (s-1)$. We use the expansions of the Airy functions in terms of Bessel functions $J_{+{1 \over 3}}$ and $J_{-{1 \over 3}}$ for $s > 0$ \cite{Abramowitz} and determine the particular solution satisfying the boundary conditions and find that
the critical buckling load $\alpha_{c}$ is given by the implicit equation
$
J_{-{1 \over 3}} \left({2 \over 3} \alpha^{1 \over 2}\right) = 0$.
Thus there exists a non-trivial small deformation solution.

For the  active filament with  boundary conditions $\theta(0) = 0$, ${\theta'} (1) =0$ and ${\theta''} (1) = 0$, we again seek solutions in terms of the Airy functions. Recasting the equation for $\theta$ as
$
 \theta = \int_{0}^{s} Z(s') \:ds'$,
and defining $\varphi \equiv \beta^{1 \over 3}(s-1)$, we obtain the differential equation
\begin{equation}
 {Z_{\varphi \varphi}} - \varphi Z =0, \:\:\:Z(0) = 0, \:\:\:{Z_{\varphi}}(0) =0.
\end{equation}
Solutions of the form
$ Z = C_1 {\mathcal{A}}(\varphi) + C_{2} {\mathcal{B}}(\varphi) $ can satisfy the required boundary conditions only when
$ C_1 {\mathcal{A}}(0) + C_{2} {\mathcal{B}}(0) = 0$,  and $C_1 {\mathcal{A}}'(0) + C_{2} {\mathcal{B}}'(0) = 0$ are satisfied simultaneously. This is
possible only when $C_1 = 0$ and $C_2=0$.
One can have small deformations solutions that are unsteady or large deformation steady solutions (the latter being ruled out by our simulations) .

\section{Coarse grained continuum model for $ \ka^{-1} \rightarrow 0$}

\begin{figure}
\begin{center}
\includegraphics[width=8cm]{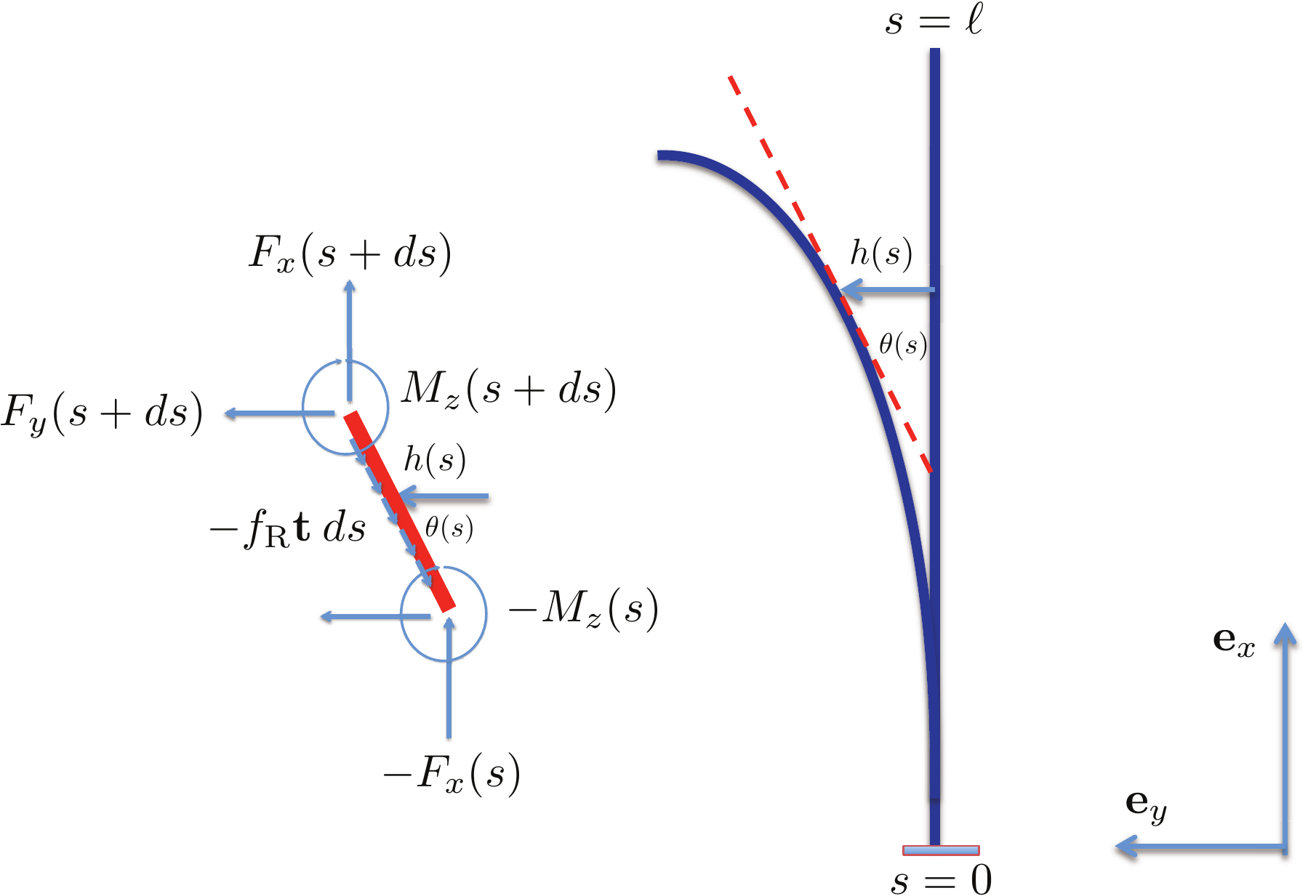}
\caption{ Schematic of the bent state and reference sketch showing the force and torques acting on a elemental slice. }
\end{center}
\end{figure}

While the previous analysis suggests that small deformation static shapes are unstable, unsteady (oscillatory) or unstable solutions can still exist. To derive the equations governing unsteady  shapes that may arise, we
begin by choosing a convenient parametrization for the filament shape.
Figure 1 is a schematic of the geometry and also shows the  free-body diagram of the forces and torques acting on an elemental length of the filament. We coarse-grain the discrete filament of attached self-propelled Brownian particles into a continuous inextensible filament of length $\ell$ and diameter $\sigma$ moving in the $x-y$ plane. The active compressive forces are also coarse-grained into a force density $f_{p}$ acting anti-parallel to the tangent vector.  Choosing the arc-length $s$ as our variable, we locate the origin of our stationary co-ordinate system at the clamped base $s=0$. The other end $s=\ell$ is free to move in the $x-y$ plane. The beating is characterized by the sequence of shapes generated as a test material point at $s$ with cartesian coordinates $(x_{p}(s), y_{p}(s))$ moves in the Newtonian liquid of viscosity $\mu$. In the long aspect ratio limit $\ell/\sigma \gg 1$, the angle $\theta(s,t)$ made by the  centerline of the inextensible filament with the $x$ axis serves as a convenient indicator of the filament shape. With this parametrization, we can express the dual vectors tangent, ${\bf t}(s,t)$  and normal ${\bf n}(s,t)$ to the centerline solely in terms of $\theta(s,t)$. Note that at $(x_{p}, y_{p})$, the increments along the cartesian directions are related to $ds$ by, $dx = \cos{\theta}\:ds$ and $dy = \sin{\theta}\:ds$.

We now work in the small deformation limit
$\sin{\theta} \approx \theta$, and only consider forces that generate  curvatures that are much greater than $\sigma$. Referring to the free body diagram in Figure (1), we note that in the physically relevant non-inertial (small Reynolds number) limit, the viscous forces, elastic  and active forces are all in balance. To make progress we decompose the total force resultant at a cross-section $s$, ${\bf F}(s,t)$ into its tangential $T$ and normal $N$ components
\begin{equation}
{\bf F} = T {\bf t} + N {\bf n}.
\end{equation}
Furthermore, we assume that the viscous forces are given by the values obtained from local resistivity theory \cite{Resistivity} - thus the viscous force per unit length at $s$ is
\begin{equation}
{\bf f}_{v} = - (\eta_{\|}u_{\|}{\bf t} + \eta_{\perp}u_{\perp}{\bf n})
\end{equation}
where $\eta_{\|}$ is the effective viscous resistance per unit length for motion of the filament along the tangent, $\eta_{\perp}$ is the resistance per unit length for motion along the local normal, $u_{\|}$ is local velocity of the centerline along the tangent vector and $u_{\perp}$ is the filament velocity along the normal. The force balance then is
\begin{equation}
-F(s) + F(s+ds) -f_{p}{\bf t} \:ds = -{\bf f}_{v}.
\end{equation}
Using
\[
{{\bf F}'} = \left({T'} - N {{\theta}'}\right) {\bf t}
+  \left({{N}'} + T {{\theta}'}\right) {\bf n}
\]
we balance force components to obtain the equations
$
\left({{N}'} + T {{\theta}'}\right)  = \eta_{\perp}u_{\perp}
$ and $
\left({{T}'} - N {{\theta}'}\right) = \eta_{\|}u_{\|} + f_{p}.
$
A balance of moments acting on the differential element yields
\begin{equation}
{{{\bf M}}'} + {\bf t} \times (N {\bf n} + T {\bf t} ) = 0
\end{equation}
Combining all the above expressions, we can eliminate $N$ and relate the bending moment per unit length to the angle $\theta$ using
${M} = {\kappa} {{\theta}'}$ to finally obtain the coupled non-linear equations for the tension, $T$ and angle $\theta$
 \begin{equation}
-\kappa  {{\theta}'''} + T {{\theta}'}  = \eta_{\perp}u_{\perp}
\end{equation}
\begin{equation}
{{T}'} + \kappa { \theta''} {{\theta}'} - f_{p}   = \eta_{\|}u_{\|}
\end{equation}

To close these equations, we need to relate the velocity of the filament to its shape and properties. For an
inextensible filament, we have following the steps in \cite{Resistivity}
\[
{ds \over dt} = 0, \:\:\:\:{\mathrm{and}}\:\:\:\:{d{\bf t} \over dt} = {d{\theta} \over dt}{\bf n} = {\partial{\theta} \over \partial{t}} {\bf n}
\]
and at the same time,
\[
{d{\bf t} \over dt} = \left({u'_{\|}} - {\theta'} u_{\perp}\right) {\bf t} +  \left({u'_{\perp}} + {\theta'} u_{\|}\right) {\bf n}.
\]
Using these expressions to eliminate the filament velocities in favor of the filament shape and shape changes and indicating derivatives with respect to time as subscripts,
we find
\begin{equation}
{{\theta}_{t}} = {{{u}}'_{\perp}} + u_{\|}{{\theta}'}, \:\:\:{\mathrm{and}}\:\:\:
0 = {{{u}}'_{\|}} - u_{\perp}{{\theta}'},
\end{equation}
and consequently
\begin{equation}
T'' = - \kappa (\theta'' \theta')' + (f_{p})' + {\eta_{\|} \over \eta_{\perp}}\: \theta' \:
\left( -\kappa \theta''' + T\theta'\right)
\end{equation}
and
\begin{equation}
\eta_{\perp} \theta_{t} = -\kappa \theta'''' + (T \theta')' +  {\eta_{\perp} \over \eta_{\|}}\:\theta'\:(T'
+ \kappa \theta'' \theta' - f_{p}).
\end{equation}
A full numerical solution to these equations under the constraint $\eta_{\perp} = \eta_{\|}$ will yield the pre-factors in equations (2) and (3) of the main text.

\section{A first passage time calculation for finite $\ka$}

We adopt a simplified picture of the the local dynamics and fluctuating forces generated by the self-propelled spheres. Since our initial
condition is a vertical filament, this special orientation results
in a critical load for buckling before any sustained bending can occur.

Let the total active force $f_{p}(s)$ be separated into a quasi-static average part
and a rapidly fluctuating part: $f_{p} (s) = \langle f_{p}(s) \rangle + \hat{f}_{p}(s)$. Fluctuations arise as the orientations of
individual active elements  escape from a favorable configuration (anti-parallel to the tangent) to an unfavorable  configuration where the active force contribution is negligible - chosen to be along the normal, ${\bf n}(s)$.

In the diffusion (noise) dominated limit
$\ka \ll 1$, the active, polar spheres are very animated, re-orienting themselves constantly with respect to their neighbors and over very small time scales compared to the response of the filament. Thus, we then have both $\ka \ll 1$ and the characteristic filament beating timescale $T_\text{f} \sim \eta f^{-{4 \over 3}}_{p} \kappa^{{1 \over 3}} \gg D^{-1}_{R}$. Irrespective of the magnitude of the constant part of $f_\text{p}$
the filament does not undergo correlated beating in this limit. In agreement with this statement, the simulations show only irregular beating or no beating for $\ka<1$ (Figure 4 of the main text).

Consider now the case when $ 1 \ll \ka < \infty$. Simulations (Figure 4 of main paper) demonstrate three distinct regimes. For low propulsion forces below the critical value $\fp < \fc$, we do not  observe  any statistically significant beating (green triangles in Figure 4 of the main text) - the straight configuration remains stable and variations in $\langle \Lee\rangle\simeq \ell$ due to fluctuations in the directions
of $\vect{p}_{i}$ are insignificant. As $\fp$ increases at constant $\ka$, we go from the no-beating regime (A) to an intermediate irregular beating regime (B) and finally to the regular beating regime (C). Note that the critical active force density that separates both transitions (A $\rightarrow$ B as well as B $\rightarrow$ C) depends on the passive filament elasticity ($\ell$ and $\kappa$) and also on $\ka$.

To understand the physical mechanisms underlying these transitions, we first note that the scaling for $\fc$ derived in the main text (Eqn. 2) corresponds to the noise-less regime  $\ka^{-1} =  0$. In the weak noise limit,  the effective time to escape is extremely large due to the high barrier to rotation of polarity vectors. Thus for $\fp < \fc$, the filament is straight and for $\fp > \fp$ the filament first buckles and then beats periodically  with a frequency that scales as $\fp^{4/3}$.

With this in mind, consider fixing the active density $\fp$ to a value greater than $\fc$ and varying $\ka$ (moving along the horizontal axis in the figures).
As  $\ka$ is increased from unity, the enhanced correlation of particle propulsion directions results in two trends - evident both for the clamped (Fig 4a) and pivoted (Fig 4b) cases. First, at some value of $\ka$, the force density becomes large enough to destabilize the straight configuration (regime A)  resulting in a regime (B) characterized by erratic beating. This $\ka$-dependent critical density decreases as $\ka$ increases,
saturating to the noiseless continuum value as  $\kappa \rightarrow \infty$. In regime (B), the mean active force density is large enough to cause buckling over the entire length of the filament $\ell$ and the orientation potential well has a depth (proportional to $\ka$) which is large enough that few polar spheres reorient. Thus, with increasing $\ka$ {\em there is
increasing correlation between the sphere self-propulsion directions
as well as an increasing time over which the correlations are sustained}. Since the self-propulsion directions tend to point along the filament tangent, there is   a typical time, $T_\text{c}$, for which the propulsion direction remains along the local tangent vector. The last trend is particularly relevant to whether or not there is sustained, regular, oscillatory beating. In particular, we hypothesize that irregular beating (regime B) arises whenever the correlation time  $T_\text{c}$ of  propulsion directions is shorter than the filament oscillatory timescale in the noiseless, perfectly aligned limit, i.e. $T_\text{f} \sim \eta_{\perp} (\kappa /\fp^4)^{1 \over 3}$.

In summary, irregular beating arises when $T_\text{c} \ll T_\text{B}$ while regular beating (regime C) arises when $T_\text{B} \gg T_{c}$. Thus we expect the critical curve characterizing the B $\rightarrow$ C transition to satisfy  $T_\text{f} \sim T_\text{c}$.

The final part of the analysis requires estimating this correlation time $T_\text{c}$.
We first rewrite the harmonic angular potential, $U_\text{a} =\frac{\ka}{2} ( \vect{p}_i- \vect{b}_i)^2$ (which biases the polar spheres to orient anti-parallel to the tangent), in terms of the angle $\zeta$. Thus, $U_{a} = \ka (1-\cos{\zeta})$.  We recognize that $T_\text{c}$  also corresponds to the mean passage time for a critical number of self-propulsion directions within a local patch of spheres to escape from a perfectly aligned configuration, in which they drive filament compression, to a stable non-compressive configuration.
For the calculation we choose the  latter  to be a configuration in which the polarity vector and the local tangent are in the same direction ($\zeta = \pi$) so that the forces are extensional and thus no compressive instabilities are possible.
\begin{figure}
\begin{center}
\includegraphics[width=\columnwidth]{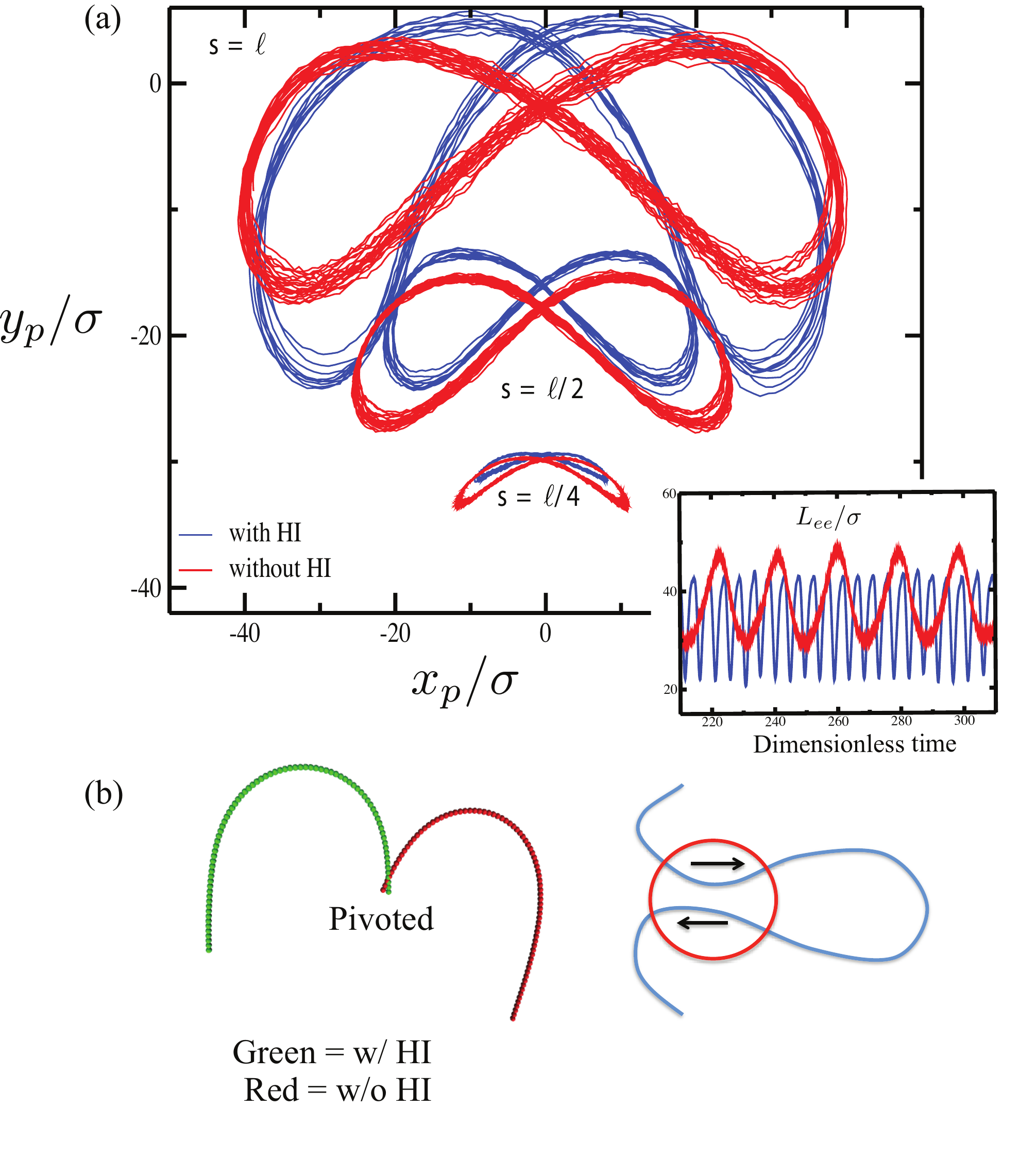}
\caption{(a) Trajectories of material points on the filament with a clamped boundary condition, with (blue) and without (red) full hydrodynamic interactions (HI).  Inset: the end-to-end distance $L_\text{ee}$ of beating filaments as a function of time. (b) The configurations of a filament with a pivoting boundary condition for simulations with (green) and without (red) full HI. For (a) and (b) the filament length is $\ell=80$, and the active force density is $f_\text{p}=10$. (c) Schematic of a filament configuration where relative motions of adjacent filaments encounter increased viscous drag when full HI are considered.}
\label{SI_fig2}
\end{center}
\end{figure}

The energy difference between the two configurations $\zeta = 0$ and $\zeta = \pi$ respectively is  $2\ka$. For $\ka > 1$, equilibration in $\zeta$ may be assumed, with the polar spheres crossing the barrier diffusively. In the over-damped limit, this flux can be related to a frequency of crossover or the disrupting frequency. While a complete calculation of the escape frequency is complicated, an asymptotically accurate estimate obtained using the steepest descent method and valid in the over-damped, viscosity controlled limit gives \cite{Hanggi}:
\[
\Gamma  \sim {{\omega_{0} \omega_\text{b}} \over {2 \pi \gamma_\text{b}}} e^{-2\ka}
\]
where $\gamma_\text{b}$ depends on the resistivity in the vicinity of the escape point ($\zeta = \pi$),
the frequency of the stable basin is given by $\omega_{0}  \sim |U''_{a}(0)|^{1 \over 2} \sim \sqrt{\ka}$ and the frequency of the barrier is given by
$\omega_\text{b} \sim |U''_\text{a}(\pi)|^{1 \over 2} \sim \sqrt{\ka}$.
Since the diffusivity and thus the viscosity are constant in all our  simulations, we ignore multiplicative factors.
Setting this frequency to be of the same order as the zero-noise frequency (Equation 3 in the main text)  we obtain a preliminary estimate
for the critical line of $\fp \sim \ka^{3 \over 4} (e^{-2\ka})^{3 \over 4}$.

This estimate needs to be modified as follows.  First,  a non-zero critical density for buckling arises due to the non-generic initial configuration. However once buckled, the net force determines the beating and the beat frequency is a function of $\fp$, not $\fp-\fc$.
Thus we introduce $A^{*}$, an adjustable parameter which ensures that the critical force tends to the noiseless continuum limit as $\ka \rightarrow \infty$.
Second,
the irregular beating behavior requires that a critical number of motors escape from the well, which we account for by introducing a pre-factor $B^{*}$  as a free parameter.  Satisfying these conditions leads to the estimate
\begin{equation}
\fp  \sim A^{*} + B^{*} \: \ka^{3 \over 4} (e^{-2\ka})^{3 \over 4}.
\end{equation}
We rewrite this in dimensionless form by redefining $A \equiv A^{*}\ell^{3}/\kappa$ and $B \equiv \ell^{3}/\kappa$ to obtain 
 equation 4 of the main text.
\begin{equation}
{\fp \ell^3 \over \kappa}  \sim A + B \: \ka^{3 \over 4} (e^{-2\ka})^{3 \over 4},
\end{equation}
\section{Effects of hydrodynamic interactions}

In simulations that were discussed in the main text, we used a freely draining model for polymer, where anisotropy of drag on the filament as well as
hydrodynamic interactions between different parts of the filament due to the flow set up in the surrounding fluid were neglected.

In order to capture the effect of full hydrodynamic interactions (HI) on the dynamics of attached filaments in the limit  $\ka \rightarrow \infty$, we used a hybrid simulation technique, in which molecular dynamics simulations for the filament were combined with a mesoscale hydrodynamic simulation method called  multi-particle collision dynamics (MPC) for solvent~\cite{mpc}.  In this approach we model the solvent as a collection of $N$ point-like particles of mass $m$,  whose velocities are determined by a stochastic process. Two steps are performed at each time point to evolve a trajectory. In the streaming step, the particles move ballistically for a time interval $h$ that may be understood as a mean collision time. In the second step (the collision step), the particles are sorted to cells of a square lattice with lattice constant $a$, and the particle velocities relative to the center-of-mass velocities of the cell, are rotated by an angle $\alpha$. The direction of rotation is chosen randomly. The conservation of mass and momentum in each cell yields accurate long range HI. The dynamics of the active filament is meanwhile simulated using a standard velocity-Verlet algorithm, and the filament-fluid interaction is implemented by including the filament monomers in the collision step and allowing for the appropriate momentum transfer.

Simulation parameters for the solvent were $\alpha = 130^\circ$, $h=\sqrt{ma^2/k_BT}$, and the mean number of particles per cell $\langle N \rangle =20$. For the filament, the monomer mass $M=m\langle N \rangle$, the bond length was $b=a$, and the monomer diameter was $\sigma=b$. A constant temperature was maintained by local rescaling of the solvent velocity.

We next compared the dynamical behavior of both clamped and pivoting filaments with and without full HI. For the clamped boundary condition,  we found the same critical active force density for buckling $f_\text{c}$ as for the simplified local hydrodynamics model, and for $f_\text{p} > f_\text{c}$ the filament interacting via full HI displays the same flagella-like beating as does the simplified model.  Furthermore the expression for beating frequency as a function of active force density $f_\text{p}$ is the same in both cases. A detailed comparison of the beating patterns of the two predictions is made by tracking specific material points along the filament contour. We plot in Figure 2(a) the trajectories of the three material points located at $s=\ell/4, \:\ell/2$ and $\ell$ when $\ell = 80$. The results demonstrate that beating patterns in the presence and absence of full HI are qualitatively similar, with the model filament transcribing a figure-of-eight in both cases.
We see that including full HI leads to slightly smaller lateral amplitudes (Fig. 2(a)) and increases the beating frequency (inset).    The reduction in both relative motions between filament parts as well as a reduction in beating amplitude (due to the increased viscous interactions) when combined with the constant energy input due to activity may account for the increase in frequency. However, we reiterate that the differences in beating patterns with and without HI are not qualitative and that expressions for the frequency in terms of the active force density, $\fp$ and the critical force for buckling are the same.  A pivoting boundary condition with full HI yields a rotating filament with similar shapes as found for the simplified model (Figure 2(b)).

Non-local hydrodynamics play a more important role for very long filaments, where parts of the filament separated by distances much greater than the filament thickness $\sigma$ come close to each other (Figure 2c). Our minimal simplified model incorporates self-avoidance accurately via the excluded volume interaction force,  and isotropic hydrodynamic resistivity tensors to compute the local viscous drag. A slightly more accurate  expression for the drag per unit length
${\bf f}_{v}(s)$ at material point ${\bf x}$ with velocity ${\bf x}_t$ that incorporates the effect of anisotropy (thus the filament is no longer a freely draining polymeric filament) is given by
\[
{\bf f}_{v}(s,t) = -  {{2 \pi \mu} \over \ln{(\ell/\sigma)}} (2 \bdelt - {\bf t}{\bf t})\bcdot{\bf x}_{t}.
\]
This expression however still neglects the full non-local hydrodynamics that arises naturally from our hybrid MPC simulation. While the qualitative response (beating or rotating) is expected to be the same for long filaments,  the increased role of viscous forces when the filaments are very close to one another (as in highly looped, beating configurations or tightly wound rotating shapes) modifies the beating patterns significantly. A systematic analysis of the changes in beating frequencies and shapes of long, self interacting filaments, due to fluid mediated interactions is a topic for our future study.